\documentclass[sigconf]{acmart}
 
\copyrightyear{2026}
\acmYear{2026}
\setcopyright{cc}
\setcctype{by}
\acmConference[CIKM '26] {Proceedings of the 35th ACM International Conference on Information and Knowledge Management}{November 7--11, 2026}{Rome, Italy.}
\acmBooktitle{Proceedings of the 35th ACM International Conference on Information and Knowledge Management (CIKM '26), November 7--11, 2026, Rome, Italy}
\acmISBN{979-8-4007-2539-5/2026/11}
\acmDOI{10.1145/3799682.3839274}
 
\usepackage{booktabs}
\usepackage{enumerate}
\usepackage[shortlabels]{enumitem}
\setlist[enumerate,1]{topsep=3pt,itemsep=2pt,parsep=0pt}
\setlist[enumerate,2]{topsep=1pt,itemsep=1pt,parsep=0pt,leftmargin=1.5em}
\usepackage{microtype}
\usepackage{multirow}
\usepackage{enumitem}
\usepackage{makecell}
\usepackage{balance}
\setlist[itemize]{beginpenalty=0, midpenalty=0, endpenalty=0}

\begin{document}
 
\title{IDs That Mean Something: A Hands-On Tutorial on Semantic IDs for Recommender Systems}

\author{Roger Zhe Li}
\affiliation{\institution{Huawei Ireland Research Center}
             \country{Dublin, Ireland}}
\email{roger.zhe.li@huawei.com}

\author{Aghiles Salah}
\affiliation{\institution{Huawei Ireland Research Center}
\country{Dublin, Ireland}}
\email{aghiles.salah@h-partners.com}

\author{Mete Sertkan}
\affiliation{\institution{Huawei Ireland Research Center}
\country{Dublin, Ireland}}
\email{mete.sertkan@h-partners.com}

\author{Changhong Jin}
\affiliation{\institution{University College Dublin}
             \country{Dublin, Ireland}
             }
\email{changhong.jin@ucd.ie}

\author{Yingjie Niu}
\affiliation{\institution{Huawei Ireland Research Center}
            \country{Dublin, Ireland}
            }
\email{yingjie.niu1@h-partners.com}

\author{Ruihai Dong}
\affiliation{\institution{University College Dublin}
            \country{Dublin, Ireland}
            }
\email{ruihai.dong@ucd.ie}

\author{Barry Smyth}
\affiliation{\institution{University College Dublin}
             \country{Dublin, Ireland}
            }
\email{barry.smyth@ucd.ie}

\renewcommand{\shortauthors}{Roger Zhe Li et al.}
 
 
\begin{abstract}
Traditional recommender systems rely on arbitrary integer IDs to identify items, which carry no semantic information and force models to learn representations from interaction data. While leveraging content information can partially mitigate cold-start, this paradigm inherently suffers from exploding embedding tables and scenario-specific architectures. Recent years have seen the emergence of semantic IDs, which encode semantic information directly into the item identity, reflecting content, hierarchy, and relationships.
This framework addresses the above limitations by offering a principled way to encode large-scale semantic information into item identity, benefiting both generative and discriminative applications and unlocking more native unification across modalities, domains, and tasks.
This tutorial traces the evolution from raw IDs to semantic IDs as a scalable means of information utilization, offering a comprehensive review of their development and applications across multi-modality, cold-start, and cross-domain recommendation. Hands-on exercises guide attendees through building and evaluating semantic IDs, training models, and exploring cross-domain transfer. We conclude by discussing open challenges, including dynamic semantic IDs and multi-stakeholder utility. All materials of the tutorial could be found at the official website \textit{\url{https://semantic-id-tutorial.github.io}}.

\end{abstract}


 
\maketitle
 
\section{Audience, Prerequisites, and Benefits}

This tutorial targets researchers and practitioners with a general interest in recommender systems, information retrieval, and representation learning.

Attendees are expected to have basic familiarity with machine learning concepts and recommender systems fundamentals. No prior knowledge of semantic IDs or generative retrieval is required.

The topic is particularly relevant to CIKM researchers and practitioners working on recommender systems, information retrieval, knowledge representation, and scalable machine learning. 
Their foundational skills in designing efficient retrieval systems, modeling semantic relationships, and evaluating large-scale models position them well to both understand and advance semantic ID-based approaches. 
As semantic IDs are rapidly gaining traction in both academia and industry, this tutorial offers a timely opportunity to build on these strengths while getting up to speed with the state of the art. Participants will benefit from a structured overview of the field, from foundational concepts to advanced applications, as well as hands-on experience building and evaluating semantic ID-based models. 
By the end of the tutorial, attendees will

\begin{itemize}
    \item have the conceptual grounding and practical tools to apply semantic IDs in their own research or production systems.
    \item be able to identify scenarios where semantic IDs have the most impact, including cold-start and cross-domain recommendation.
    \item understand open challenges and emerging directions beyond conventional recommendation, like dynamic semantic IDs, multi-stakeholder utility, and semantic planning.
\end{itemize}

\section{Tutorial Overview} 
\label{sec:bgd_motivation}

Recommender systems have undergone a significant transformation in how they represent and identify items, evolving from raw IDs as simple lookup keys to semantic IDs as structured interfaces for utilizing and sharing information across recommendation settings.
 


\subsection{Raw IDs}
We start with raw IDs. For much of the past two decades, item identity in industrial systems was anchored in discrete and arbitrary integer identifiers (e.g., book \#892, movie \#4200). While these identifiers carry no semantic information about the items, they enabled exact lookup, logging, and item-specific memorization at scale~\cite{DBLP:conf/recsys/ChengKHSTAAICLAHJLS16}.
These raw IDs resolved non-trivial challenges in information utilization in recommendation, such as architectural stability, cross-component consistency, and high-concurrency serving
. However, they also come with fundamental limitations: raw IDs provide no identity-level transfer across semantically similar items, leaving cold-start unsupported at the identity level and embedding tables growing unboundedly with the catalog.

\subsection{Auxiliary Information}

To capture richer item semantics and mitigate cold-start, an important line of work has explored incorporating auxiliary item information into recommendation models. Often referred to as content-based or multimodal recommender systems, the core idea is to derive item representations from both user-item interactions (collaborative filtering) and item content~\cite{DBLP:journals/csur/DeldjooSCP20, DBLP:conf/recsys/TruongSL21}, which may come in the form of text,
images,
graphs,
audio,
or videos.
This auxiliary information complements collaborative signals in capturing how items are consumed together, while enabling the derivation of representations for new items at inference time, effectively addressing cold-start and bridging the gap across domains. 

Despite these undoubted benefits, significant challenges remain under the raw-ID setting: embedding tables grow prohibitively large with item vocabulary size~\cite{DBLP:conf/recsys/FukumotoSSTSMX25}, recommendation architectures become increasingly complex to accommodate various modalities~\cite
{DBLP:journals/csur/LiuHXZGWLT25}, and modality-specific designs limit the usability of new or different auxiliary information in existing model settings~\cite{DBLP:journals/internet/TruongSTGL21}.

\subsection{Semantic IDs}
While auxiliary information has significantly advanced recommender systems, it remains peripheral to item identity itself. An emerging paradigm shifts this boundary: rather than treating semantic information solely as an auxiliary input, semantic IDs encode part of that information directly into the item identifier. 
Concretely, items are represented as discrete token sequences, typically obtained through vector quantization or clustering methods, such as RQ-VAE and RQ-KMeans~\cite{DBLP:conf/nips/RajputMSKVHHT0S23, 
DBLP:conf/cikm/LuoCSYHYLZ0HQZZ25}. This turns item identity into a structured, model-facing object that can reflect content, hierarchy, relationships, and behavioral signals, enabling cross-item generalization while preserving the lookup and memorization properties that made raw IDs indispensable in practice.

In addition to reviewing the main methods for deriving semantic IDs in the recommendation context, we discuss the key design choices behind semantic ID construction, including what information is quantized, how token vocabularies are organized, and how to balance compactness, stability, semantic coherence, and downstream usability.

\subsection{Semantic IDs in Recommender Systems}
Originally developed for computer vision and approximate nearest neighbor search, semantic IDs have since been increasingly adopted in recommender systems. Their most prominent application is perhaps Generative recommendation, where the model learns to generate the semantic ID of the next item, a direction that is extensively explored, establishing semantic IDs as a core building block of generative recommendation~\cite{DBLP:conf/nips/RajputMSKVHHT0S23}. 

Semantic IDs are also being leveraged as structured input features in discriminative ranking pipelines, with successful deployments at Google~\cite{DBLP:conf/recsys/SinghVMKSZHHWTCY24}, Kuaishou~\cite{DBLP:conf/cikm/YeSSWWJ25}, and Meta~\cite{DBLP:conf/recsys/ZhengHPRWXNL00L25} demonstrating that structured representations improve cold-start performance and embedding stability in this context as well.


We deep-dive into TIGER~\cite{DBLP:conf/nips/RajputMSKVHHT0S23} and the YouTube ranking model~\cite{DBLP:conf/recsys/SinghVMKSZHHWTCY24} as two representative use cases of semantic IDs across generative and discriminative settings. The focus is on how Semantic IDs support scalability, generalization, and mitigate cold-start.


\subsection{Multimodality and Collaborative Signals}

Semantic IDs are mainly derived from item content, which may come in various forms including text, images, audio, and video. This naturally raises the question of multimodality: \textit{how should these diverse sources of information be modeled and fused into semantic IDs?} This tutorial examines existing work along two emerging families of methods, namely early fusion and late fusion~\cite{DBLP:journals/corr/abs-2503-23333}, discussing their respective trade-offs in terms of complexity, flexibility, representational quality, and point out some potential limitations such as redundancy~\cite{DBLP:journals/corr/abs-2601-08764}.

Beyond content, recent work has also established the importance of collaborative signals, capturing how users interact with items, for recommendation accuracy. We cover the main approaches to incorporating such signals into semantic IDs, including collaborative embedding quantization~\cite{DBLP:conf/cikm/0006P0000LJ025},
content model/embedding fine-tuning~\cite{DBLP:journals/corr/abs-2601-22694, DBLP:conf/cikm/LuoCSYHYLZ0HQZZ25}, and contrastive training on collaborative data~\cite{DBLP:conf/www/HeHHKMMSTWWYBCC26}. This part highlights a central design tension: semantic IDs should be meaningful enough to reflect item content, yet recommendation-aware enough to capture user interaction patterns in practice.

\subsection{Cross-Domain Transfer} 

Cross-domain recommendation presents a challenging setting where users and items may have little or no interaction history in a target domain. Traditional approaches rely on explicit source-target alignment, introducing model complexity that becomes intractable as the number of target domains grows ~\cite{DBLP:conf/wsdm/CaoL0GL023}. We briefly review the representative methods in this context ~\cite{DBLP:conf/ijcai/YuanYB19, DBLP:conf/recsys/SalahTL21}, and highlight their remaining limitations.

We then turn to Semantic IDs, which offer a more principled alternative: shared semantic structure across domains emerges naturally from the quantization process, enabling seamless and scalable cross-domain transfer without explicit alignment. We examine emerging methods in this space~\cite{DBLP:journals/corr/abs-2507-12871, DBLP:conf/aaai/HuLW26} and show how this paradigm directly addresses the limitations of conventional approaches.

\subsection{Future Directions \& Open Challenges}

We close by discussing emerging directions. One is cross-task unification, where semantic IDs provide shared representations across recommendation and 
search~\cite{DBLP:conf/recsys/PenhaDNPTVLLHFB25, 
DBLP:conf/recsys/ShiXZZZSY25}. 
Another is semantic planning, where semantic IDs support reasoning about what the next exposure should accomplish under user, platform, and provider objectives. We also discuss open challenges including ID stability under catalog dynamics and evaluation beyond hit-rate 
metrics.

\subsection{Hands-on Sessions}
This is a hands-on tutorial, whereby lectures are complemented by two practical sessions. In the first, attendees train RQ-KMeans to derive semantic IDs, inspect the resulting token structure, and compare it against a raw ID-based baseline on a generative recommendation task. 
In the second, attendees investigate the architecture proposed in~\cite{DBLP:journals/corr/abs-2507-12871}, leveraging semantic IDs in a multi-domain setting, and assess their cross-domain transfer capabilities. 

Both sessions use the Amazon Reviews dataset~\cite{DBLP:conf/sigir/McAuleyTSH15}, with all instructions and code provided as self-contained Jupyter notebooks that run on free-tier resources such as Google Colab. To keep the sessions focused on semantic IDs, we leverage Cornac's~\cite{DBLP:journals/jmlr/SalahTL20} generative recommendation model implementations, and pre-compute content-based item embeddings and baseline models when necessary. To allow participants to prepare in advance, the materials are made publicly available at \url{https://github.com/saghiles/semantic-id-tutorial}.

\section{Outline \& Equipment Requirements}
\label{sec:Outline}
The half-day tutorial consists of 3 hours of core content, plus two coffee breaks, organized as follows:
\begin{enumerate}
    \item Introduction to recommender systems, raw item IDs, and auxiliary content  (15 min)
    \item Semantic IDs: introduction \& overview (15 min) 
    \item Leveraging semantic IDs in recommender systems  (20 min)
    \item[] \textit{--- Coffee break --- (15min)}
    \item Hands-on: building and using semantic IDs (25 min)
    \item Multimodality and collaborative signals (25 min)
    \item Cold-start mitigation and cross-domain transfer (25 min)
    \item[] \textit{--- Coffee break --- (15min)}
    \item Hands-on: cross-domain usage of Semantic IDs (25 min)
    \item Future directions and open challenges (15min)
    \item Q\&A and open discussion (15 min)
\end{enumerate}

For presentation, we require standard equipment: an LCD projector, at least one projection screen, and microphones. Attendees are expected to bring their own laptops for the hands-on sessions.

\section{Related Tutorials}
Several recent tutorials cover adjacent content but differ in focus. Table ~\ref{tab:related} provides a brief summary of the most closely related ones. 

\begin{table}[h]
    \caption{Comparison with Related tutorials.}
    \label{tab:related}
    \footnotesize
    \renewcommand{\arraystretch}{0.9}
    \begin{tabular}{@{}lp{3cm}p{3.6cm}@{}}
        \toprule
        Venue & Title & Core angle \\
        \midrule
        CIKM '25 & Towards Large Generative Recommendation: A Tokenization Perspective~\cite{DBLP:conf/cikm/Hou0SWWCM25} 
            & How tokenization choices shape generative recommendation. \\
        WWW '25 & Generative Recommendation Models: Progress and Directions~\cite{DBLP:conf/www/HouZS0WCM25} 
            & Scaling and progress of generative recommendation models \\
        SIGIR '24 & LLMs for Recommendation: Past, Present, and Future~\cite{DBLP:journals/tois/LinDXLCZLWLZGYTZ25} 
            & How LLMs are adapted for recommendation \\
        SIGIR '25 & Navigating Large Language Models for Recommendation~\cite{DBLP:conf/sigir/0001BZ00F25} 
            & LLM technical stack from architecture to deployment \\
        WSDM '25 & Recommendation with Generative Models~\cite{DBLP:conf/wsdm/DeldjooHMKSRVSK25} 
            & VAE, GAN, diffusion, and LLM applied to recommendation \\
        \midrule
        \makecell[tl]{CIKM '26 \\ (ours)} & \textbf{IDs That Mean Something: A Hands-On Tutorial on Semantic IDs for Recommender Systems} [this work] 
            & \textbf{How item identity encodes information, from raw IDs to semantic IDs, across generative and discriminative applications} \\
        \bottomrule
    \end{tabular}
\end{table}


Related tutorials can be roughly grouped into three categories: those surveying tokenization methods within the generative paradigm; those treating semantic IDs as a sub-topic within the broader scope of LLM architecture for recommender systems~\cite{DBLP:journals/tois/LinDXLCZLWLZGYTZ25, DBLP:conf/sigir/0001BZ00F25}; and those covering generative model families beyond LLMs (e.g., VAE, GAN, diffusion models) applied to recommendation~\cite{DBLP:conf/wsdm/DeldjooHMKSRVSK25}. While our discussion of semantic ID construction inevitably shares some terminology and methodology with these tutorials (e.g., RQ-VAE), our tutorial differs from all of them along three key dimensions:

\begin{itemize}
    \item \textbf{Semantic IDs as the primary focus.} We organize the tutorial around how recommender systems utilize information, tracing the evolution from raw IDs to semantic IDs. Unlike prior tutorials, we treat each stage as substantive content and cover semantic IDs across both generative- and discriminative-oriented applications~\cite{DBLP:conf/recsys/SinghVMKSZHHWTCY24,
DBLP:conf/recsys/ZhengHPRWXNL00L25,DBLP:conf/cikm/YeSSWWJ25}.
    \item \textbf{Evolution of semantic ID representations.} We cover the ongoing enrichment of semantic IDs beyond content-only signals towards representations that unify modalities, domains, and tasks. This trajectory is attracting rapid R\&D attention but has not been addressed in prior tutorials.
    \item \textbf{Hands-on exercises.} We include hands-on exercises covering semantic ID construction, model training, and the downstream use cases of semantic IDs. To the best of our knowledge, this is the first tutorial on this topic to do so.
\end{itemize}

Overall, existing tutorials have primarily focused on generative models, LLM architectures, or tokenization as a method class. Our tutorial complements these by covering how semantic IDs emerge and enable more scalable information utilization in recommender systems, across both generative and discriminative applications.

\section{Presenters}
\begin{itemize}
    \item \textbf{Roger Zhe Li\footnote{\url{https://www.zhe-li.me/}}} is a Senior Research Scientist at the Huawei Ireland Research Center,  leading ads recommendation R\&D covering multi-modal semantic IDs and CVR prediction, with the system serving over 250 million users globally. He holds a PhD from TU Delft, with publications at SIGIR, WSDM, RecSys, and ICTIR. He was recognized as an outstanding PC member at RecSys '25, and served as an invited industry speaker on multi-modal semantic IDs at ECIR '26.
    \item \textbf{Aghiles Salah\footnote{\url{http://saghiles.github.io/}}} is a Principal Research Scientist at the Huawei Ireland Research Center. His research spans machine learning and recommender systems, with publications in tier-1 conferences and journals. He co-delivered hands-on tutorials on multimodal recommender systems at RecSys '21 and WebConf '23, and is the pioneer and one of the main contributors of the Cornac recommendation framework. Over more than a decade, Aghiles taught undergraduate and master's courses on machine learning and recommender systems at the University of Paris, France.
    \item \textbf{Mete Sertkan\footnote{\url{https://metesertkan.github.io/}}} is a Senior Research Scientist at the Huawei Ireland Research Center working on ads recommendation. His research interests span recommender systems, user modeling, information retrieval, and evaluation, with publications in top-tier venues including ACL, CIKM, UMAP, RecSys, and SIGIR. He holds a PhD from TU Wien, where he taught master's courses on recommender systems, social network analysis, and innovation.
    \item \textbf{Changhong Jin} is an Assistant Professor of Computer Science at University College Dublin. His recent work covers recommender-system auditing, ad recommendation under limited availability of personalized features, and generative recommendation with semantic IDs.  
    He has contributed to publications at CIKM, The Web Conference, and ACL, and delivered a one-hour tutorial on recommender-system auditing at the ECML-PKDD 2025 AIDEM workshop.
    \item \textbf{Yingjie Niu} is a Research Scientist at the Huawei Ireland Research Center, working on Ads recommendation and multi-modal semantic IDs. His research interests span recommender systems and large language model, with publications in tier-1 venues including ACL, WWW. He holds a PhD from University College Dublin, where he taught master's courses on machine learning and information system. 
    \item \textbf{Ruihai Dong\footnote{\url{https://people.ucd.ie/ruihai.dong/}}} is an Associate Professor of Computer Science at University College Dublin. His research focuses on Machine Learning and Deep Learning with applications in recommender systems and finance. He has published widely in top venues (e.g., WWW, RecSys, ACL, IJCAI) and served on their program committees. He co-authored the RecSys '23 Best Student Paper and has co-organized the EARL Workshop series on Evaluating and Applying Recommender Systems with Large Language Models colocated with RecSys since 2024.
    \item \textbf{Barry Smyth\footnote{\url{https://people.ucd.ie/barry.smyth}}} is a Professor of Computer Science at University College Dublin and a founding figure of recommender systems, with over three decades of work on user modeling, case-based reasoning, and personalization, amassing over 28,000 citations and an h-index of 85.
    

\end{itemize}

\begin{acks}
The authors are grateful to the anonymous reviewers for their constructive comments. Special thanks go to Xingsheng Guo and Huifeng Guo from Huawei for their extensive discussions and valuable support throughout the submission process.
\end{acks}
\section*{GenAI Usage Disclosure}
Generative AI (Claude, Sonnet 4.7) was used during the preparation of this proposal to assist with language editing and sentence-level revision and rephrasing. All technical content, arguments, claims, citations, and the overall scope and framing of the tutorial were proposed, discussed, verified, and approved by the authors.

\renewcommand*{\bibfont}{\footnotesize}
\setlength{\bibsep}{1.5pt}
\bibliographystyle{ACM-Reference-Format}
\balance
\bibliography{sample-base}

\newpage

\end{document}